\documentclass[a4paper,11pt]{article}
\usepackage{jheppub}
\usepackage{amsmath}
\usepackage{graphicx}
\usepackage{subfigure}
\usepackage{amssymb}
\usepackage{xcolor}
\usepackage{multirow}
\usepackage{cancel}
\usepackage{color}
\usepackage{epsfig}
\usepackage{ulem}
\usepackage{bm}
\usepackage{listings}
\usepackage{float}
\usepackage{slashed}

\usepackage{tikz}
\usepackage{pgffor}							

\newcommand{\be}{\begin{equation}}
\newcommand{\ee}{\end{equation}}
\newcommand{\beq}{\begin{equation}}
\newcommand{\eeq}{\end{equation}}
\newcommand{\bea}{\begin{eqnarray}}
\newcommand{\eea}{\end{eqnarray}}
\newcommand{\besp}{\begin{equation}\begin{split}}
\newcommand{\eesp}{\end{split}\end{equation}}

\newcommand{\Dfbd}{\mathord{\buildrel{\lower3pt\hbox{$\scriptscriptstyle\leftrightarrow$}}\over {D}_{\mu}}}

\def\0{\textbf{0}}
\def\1{\textbf{1}}
\def\2{\textbf{2}}
\def\3{\textbf{3}}
\def\4{\textbf{4}}
\def\5{\textbf{5}}
\def\6{\textbf{6}}
\def\7{\textbf{7}}
\def\8{\textbf{8}}
\def\9{\textbf{9}}

\begin{document}

\title{Revisiting Puffy Dark Matter with Novel Insights: Partial Wave Analysis}


\author[1]{Wenyu Wang}

\author[1]{,~Wu-Long Xu}

\author[2,3]{,~Jin Min Yang}
	
\author[4]{,~Bin Zhu}

\affiliation[1]{Faculty of Science, Beijing University of Technology, Beijing, China}
\affiliation[2]{CAS Key Laboratory of Theoretical Physics, Institute of Theoretical Physics, Chinese Academy of Sciences, Beijing 100190, P. R. China}
\affiliation[3]{ School of Physical Sciences, University of Chinese Academy of Sciences,  Beijing 100049, P. R. China}
\affiliation[4]{Department of Physics, Yantai University, Yantai 264005,  P. R. China}

\emailAdd{wywang@bjut.edu.cn}
\emailAdd{wlxu@emails.bjut.edu.cn}
\emailAdd{jmyang@itp.ac.cn}
\emailAdd{zhubin@mail.nankai.edu.cn}
\abstract{ We present a comprehensive study on the self-interaction cross-section of puffy dark matter (DM) particles, which have a significant intrinsic size compared to their Compton wavelength.
For such puffy DM self-interaction cross-section in the resonant and classical regimes, our study demonstrates the significance of the Yukawa potential and the necessity of partial wave analysis:  
(i) Due to the finite-size effect of puffy DM particles, the new Yukawa potential of puffy DM is found to enlarge the Born-effective regime for the  self-interaction cross-section, compared with the point-like DM;  
(ii) Our partial wave analysis shows that depending on the value of the ratio between  $R_{\chi}$ (radius of a puffy DM particle) and  $1/m_{\phi}$ (force range), the three regimes (Born-effective, resonant and classical) for puffy DM self-interaction cross-section can be very different from  the point-like DM;
(iii) We find that to solve the small-scale anomalies via self-interacting puffy DM, the Born-effective and the resonant regimes exist for dwarf galaxies, while for the cluster and Milky Way galaxy the non-Born regime is necessary.

}

\maketitle

\tableofcontents

\section{Introduction}\label{sec1}
In the past few decades, the dark matter (DM), an enigmatic but crucial component of the universe, has been a topic of intense research \cite{Bahcall:1999xn, Springel:2006vs}. Plenty of observational evidence, such as the rotation curves of galaxies and the collision of bullet galaxies, implies the existence of DM, which primarily interacts through the force of gravity \cite{Ostriker:1973uit, Randall:2008ppe}. The large-scale structure of the universe is strongly influenced by DM, and the $\Lambda \rm CDM$ model, which is consistent with the large-scale structure data, provides a framework to study the evolution of the universe \cite{Trujillo-Gomez:2010jbn}. However, this model falls short in explaining the observed small-scale structures, such as the core-cusp problem and the diversity problem \cite{Kauffmann:1993gv, Moore:1999nt, Burkert:1995yz, Salucci:2007tm, Boylan-Kolchin:2011qkt, Oman:2015xda}. 

The observed small-scale discrepancies in the standard $\Lambda$CDM have led to the investigation of self-interacting DM (SIDM) as a possible solution. The SIDM scenario, which allows for the redistribution of mass and dissipation of energy through scattering interactions, requires a self-scattering cross-section per unit mass in the range of $0.1\mathrm{cm}^2/\mathrm{g}< \sigma/m < 10\mathrm{cm}^2/\mathrm{g}$ to match the positive observations of different galaxies \cite{Tulin:2017ara,Spergel:1999mh, Tulin:2013teo,Colquhoun:2020adl,Chu:2018faw,Chu:2018fzy,Tsai:2020vpi,Tulin:2012wi, Chu:2019awd, Buckley:2009in,Wang:2014kja,Kim:2022cpu,Kim:2021bmx,Zhu:2021pad,Wang:2022lxn}. While the simplest realization of the SIDM scenario is a SIDM particle with a light mediator, this approach is constrained from several aspects, including the requirement to satisfy the relic density and the need to avoid conflict with the Big Bang Nucleosynthesis results, Cosmic Microwave Background observations, and the indirect detection experiments \cite{Garcia-Cely:2017qpx,Bringmann:2016din,Kahlhoefer:2017umn}. Some approaches have been proposed to address these challenges, including coupling DM to lighter degrees of freedom in the dark sector \cite{Bernal:2015ova}, using p-wave annihilation processes \cite{Chu:2016pew}, introducing composite SIDM candidates such as atoms, nuclei  and bound states \cite{Cline:2021itd, Laha:2013gva}, and exploring resonant SIDM from point particle DM or dark QCD DM \cite{Chu:2018fzy,Tsai:2020vpi,Kondo:2022lgg}. The self-interactions in the SIDM scenario can also involve processes that are not purely elastic, such as those involving small mass-splitting dark states or excited states \cite{Schutz:2014nka,Zhang:2016dck,Alvarez:2019nwt,Dutta:2021wbn}.

Some recent studies have highlighted the importance of the DM size effect for the SIDM, where the size of the dark matter particles can dominate the velocity dependence of the self-interacting cross-section. The study in \cite{Chu:2018faw} considered the Yukawa interaction for DM self-scattering, and found that the size effect can be significant even in the presence of a long-range force. In a previous work \cite{Wang:2021tjf}, it was shown that the interaction responsible for the size formation can lead to a characteristic velocity distribution, which differs from the conventional puffy dark matter. Moreover, it was found that even when the strong interaction effect is negligible, the only Yukawa interaction can still yield results different from \cite{Chu:2018faw}. To compute the cross-section accurately for the puffy Yukawa potential, the schr\"odinger equation must be solved using the partial wave analysis, as in \cite{Tulin:2013teo, Colquhoun:2020adl}, where DM is treated as a point particle. The size effect has implications for various regimes, including the Born, nonperturbative, quantum, and semiclassical regimes \cite{Colquhoun:2020adl, Digman:2019wdm}. Therefore, the partial wave analysis is a more precise method for studying the puffy SIDM. In this work, we perform a comprehensive partial wave analysis for the self-interaction cross-section of puffy DM particles and examine the resonant and classical scattering regimes that arise due to the long-range Yukawa potential.

This work is structured as follows. In Section II, we demonstrate the significant differences between our approach and the conventional puffy DM model. Specifically, we provide a comprehensive partial wave analysis of the newly derived puffy DM Yukawa potential. In Section III, we present the numerical results of the scattering calculations. In Section IV, we investigate the scattering of the puffy DM on small cosmological scales. Finally, in Section V, we summarize our findings and draw conclusions.

\section{Critical Differences from Conventional Models: Puffy Yukawa Potential}\label{sec2}
In this section, the new puffy Yukawa potential will be obtained and its significance for self-scattering cross-section is shown. We consider the scattering of self-interacting DM using the Yukawa potential when the dark matter is treated as a point particle, that is 
\be \label{eq1}
V(r)=\pm \frac{\alpha}{r} e^{-m_{\phi}r},
\ee
where the notation $+(-)$ is used to indicate the repulsive (attractive) nature of the force, while $m_{\phi}$ represents the mass of the mediator. Note that in the case of a scalar mediator, the force is purely attractive. The dark fine structure constant is conventionally defined as $\alpha = g_{\chi}^2/4\pi$ with $g_{\chi}$ being the self-coupling of DM. For puffy DM, the interacting potential between two puffy DM particles takes the form:
\be \label{eq2}
V(r)=\frac{1}{4\pi}\int dV_1 dV_2
\rho_{1}({\bm r}_1)\frac{e^{-m_{\phi}
|{\bm r}_1-{\bm r}_2|}}{
|{\bm r}_1-{\bm r}_2|}\rho_{2}({\bm r}_2)\, .
\ee
It is important to note that the calculation of the interacting potential assumes spherical symmetry, resulting in an isotropic potential. To study the size effects, different profiles of puffy DM, such as tophat, dipole or Gaussian, can be selected.
However, our analysis shows that the differences caused by different puffy DM models are negligible.
Therefore, in our study we adopt the tophat charge density profile, given by
\begin{equation}
    \rho(r) = \frac{3Q}{4\pi R_{\chi}^3}\theta(R_{\chi}-r),
\end{equation}
where $R_\chi$ can be considered as the radius of the puffy DM particle and $Q$ is the total charge of a puffy DM particle.
The calculation of the interacting potential is extensive and the details are presented in  Appendix~\ref{appa}. In the following, we only provide a summary of the resulting form of the potential between two puffy DM particles: 
\begin{align}\label{eq3}
V_{\rm puffy}(r) & = \begin{cases}
~\pm g(r,y) & r<2R_{\chi}\, , \\
\hspace{5cm}\  & \ \\[-6.mm]
~\pm\alpha\frac{e^{-m_{\phi}r}}{r} 
\times h\left(y\right)&  r>2R_{\chi}\,,
\end{cases}
\end{align}
where $y=R_{\chi}m_{\phi}$.  One can clearly see from $V_{\rm puffy}(r)$
in Appendix~\ref{appa} that for $r>2R_{\chi}$ the potential is enhanced by a
factor of $h(y)\sim e^{2y}/a^5$. This suggests that the behavior of
self-interacting puffy DM is significantly different from the point DM particles. Furthermore, for $r<2R_{\chi}$, the puffy potential can be expressed in a dimensionless form, 
allowing for a comparison with the Yukawa and Coulomb potentials:
 \begin{align}\label{eq33}
 V(r=R_{\chi}x) & = \begin{cases}
~\frac{\alpha}{R_{\chi}}
\frac{1}{x} & \rm Coulomb~ potential, \\
 \hspace{5cm}\  & \ \\[-6.mm]
~ \frac{\alpha}{R_{\chi}}
 \frac{e^{-yx}}{x} &  \rm Yukawa~ potential,\\
  \hspace{5cm}\  & \ \\[-6.mm]
 ~  \frac{\alpha}{R_{\chi}}
  H(x,y),& \rm Puffy~ potential, \\
 \end{cases}
 \end{align} 
 where 
 \be \label{Hp}
 \begin{split}
 H(x,y)=&3(y^4(-2+x)^3x(4+x)-6e^{-y(2+x)}(1+y+e^{2y}(-1+y))\\
 &\times (-2(1+y)+e^{yx}(2+y(2+(-2+y(-2+x))x)))\\
 &+6e^{-y}(1+y)(2(2+(y)^2(-2+x)x )\cosh(y)-4\cosh(y(-1+x)\\
 &+4y(-1+x)\sinh(y)-\sinh(y(-1+x)) )))/(16y^6x).
 \end{split}
 \ee
It may seem obvious that the divergent pole of the potential vanishes due to the size effect. However, it is important to note that the vanishing of a pole can significantly alter the solution of the differential Schr\"odinger equation. This is one of the main motivations for this work, and we discuss the differences between the potentials and the validity of the Born approximation in the following sections.

\begin{figure}[ht]
	\centering
	\includegraphics[width=6.5cm]{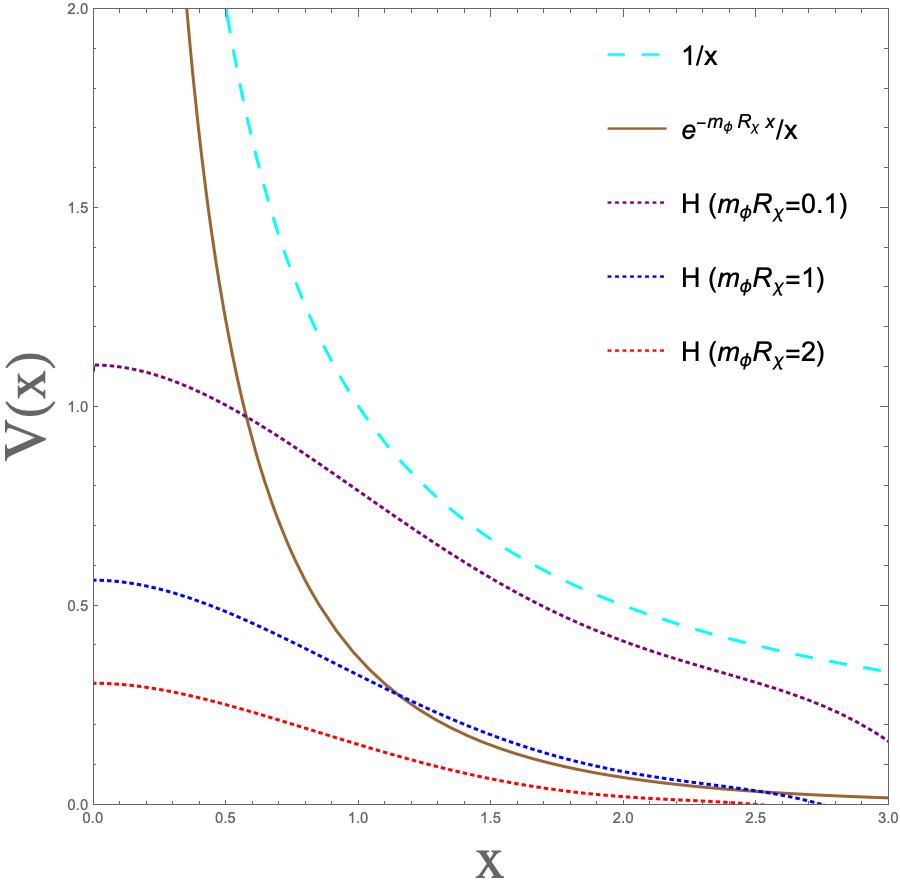}
	\includegraphics[width=8.cm]{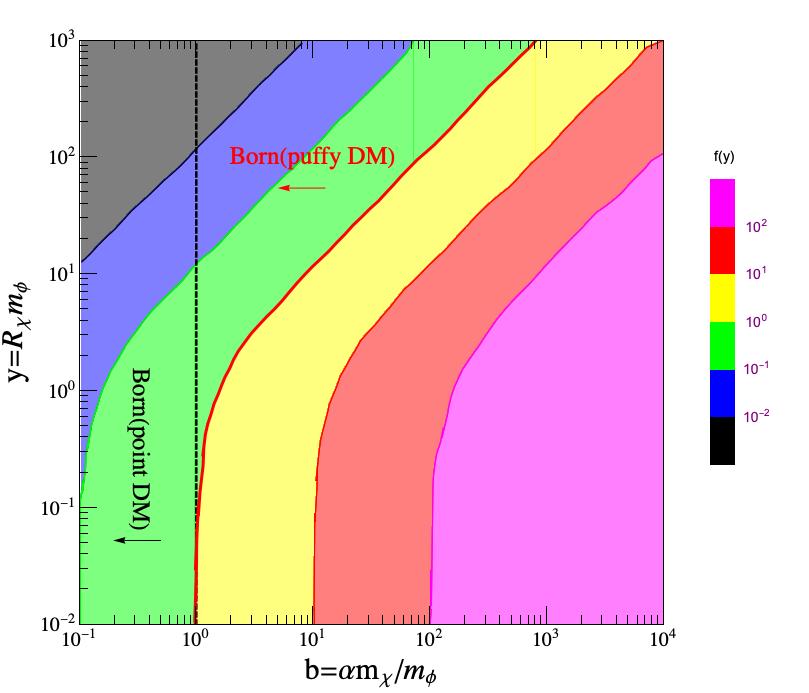}\\
	\caption{The left panel shows a comparison of the Yukawa potentials with a puffy profile, Coulomb potential, and point Yukawa potential ($y=1$) for different values of $y$ and $\alpha=1$ when $r<2R_{\chi}$. The right panel presents the parameter space of $y$ and $b$ for different values of $f(y)$ in the Born and nonperturbative regimes.}
	\label{fig1}
\end{figure}
The left panel of Fig.~\ref{fig1} demonstrates that as the ratio of radius
to force range $y=m_{\phi}R_{\chi}$ increases for $r<2R_{\chi}$, the
initial values of the potential are almost constant but continuously
decrease. In comparison to the point particle case, when $y$ is in an intermediate range,
the potential $V(r,R_{\chi} \neq 0)$ can be greater than
$V(r,R_{\chi}=0)$ within a certain range of $r$ and then drops more steeply, as shown by the brown and
purple curves in the left panel of Fig.~\ref{fig1}, due to the size effect.
The left panel of Fig.~\ref{fig1} also indicates that the Coulomb
potential serves as the upper limit for these Yukawa potentials.
Furthermore, the puffy DM potential has a 
finite smaller initial value and is
rapidly suppressed, making it negligible in solving the Schr\"odinger
equation. Therefore, if the DM particle has a size, the potential will
exhibit obvious variations. As a result, the cross-section calculated
using the Yukawa interaction should be reassessed.

In case of low energy scattering, the validity 
of the Born approximation is 
 \be\label{eqb}
 m_{\chi}\left|\int_{0}^{\infty}rV(r)dr\right|\ll1
 \, .
 \ee
For the point interacting Yukawa potential, the requirement is 
 $$\alpha m_{\chi}/m_{\phi}\ll1\, .$$ 
 In light of the puffy DM self-scattering, 
 substituting potential Eq.~\eqref{eq3} to Eq.~\eqref{eqb},
 the validity  of Born approximation
 can be derived analytically which is 
\begin{eqnarray}
m_\chi 
\left|\int_{0}^{\infty}rV(r)dr\right|
&=& \frac{ \alpha m_{\chi}}{m_{\phi}}
\frac{3(-15+y^2(15-10y+4y^3)+15(1+y^2)e^{-2y})}{10y^6}\nonumber\\
&=& f(y) \ll 1\,.
\label{pufb}
\end{eqnarray}
Here a new function $f(y)$ is defined for the estimation of the validity.
The right panel of Fig.~\ref{fig1} shows the contour map of $f(y)$ for the puffy DM,
illustrating in the parameter space of $y$ and $b=\alpha m_{\chi}/m_{\phi}$.
Note that $b$ is a dimensionless parameter which will be used to study the self-scattering in the following. We can see from this figure that  the radius effect is negligible  when $y<0.1$
and the validity of the Born approximation 
almost remains unchanged. 
 For the puffy DM self scattering the Born regime is shown as the left range of the red solid curve in the right panel of Fig.~\ref{fig1}, while the Born regime for the point particle is the region in the left of the dashed vertical line.
We see that as the value of $y$ increases, the parameter space of the puffy DM Born regime expands compared to the point particle case. For instance, for $y=10^3$, $b=100$ and $f(y)<1$, the validity of Born approximation is still met. In such an expanded Born parameter space the puffy DM self-interaction cross section can also be calculated via the field theory approach as in \cite{Chu:2018faw}.   Therefore, for puffy DM, the validity of the Born approximation depends not only on the ratio of the dark fine structure constant and the mediator mass, but also on the puffy DM charge distribution and the value of $y$. This needs to be taken into account when studying the self-interaction of puffy DM.
 
\section{Numerical Analysis of Scattering Cross-Section}\label{sec3}
 To accurately describe the self-interaction among dark matter particles, the transfer cross section $\sigma_T$ is generally adopted:
\be\label{eq4}
\sigma_T=\int d\Omega(1-\cos\theta)d\sigma/d\Omega,
\ee 
where $\theta$ is the scattering angle. 
The transfer cross section for a point-like DM particle can be classified
into different regimes based on two dimensionless parameters: 
$b=\alpha m_{\chi}/m_{\phi}$ and $m_{\chi}v/m_{\phi}$. These regimes include Born, quantum, semiclassical, nonperturbative \cite{Colquhoun:2020adl}, etc. However, for puffy DM, the approximation criteria of these regimes cannot be applied due to the piecewise nature of the Yukawa potential. Therefore, the transfer cross section $\sigma_T$ must be determined by the partial wave expansion of Schr\"odinger equation. This leads to a differential scattering cross-section value that is different from the point-like DM case for the same inputs of parameters:
\be \label{eq5}
\frac{d\sigma}{d\Omega}=\frac{1}{k^{2}}\Big|\sum_{l=0}^{\infty}(2l+1)e^{i\delta_{l}}P_{l}(\cos\theta)\sin\delta_{l}\Big|^{2}.
\ee
In this scenario, the differential scattering cross-section is given by Eq.~(\ref{eq5}), where $k$ is the momentum of the incident particle and $\delta_l$ is the phase shift of the $l$-th partial wave. 
The phase shift $\delta_l$ can be obtained by solving the Schr\"odinger equation for the radial wave function $\mathcal{R}_l(r)$ which is 
\be\label{eq6}
 \frac{1}{r^{2}}\frac{\partial}{\partial r}\left(r^{2}\frac{\partial\mathcal{R}_{l}}{\partial r}\right)+\left(E-V(r)-\frac{l(l+1)}{m_\chi r^{2}}\right)\mathcal{R}_{l}(r)=0.
\ee
The phase shift $\delta_l$ is obtained from the asymptotic solution for $\mathcal{R}_l(r)$:  
\be \label{eq7}
\underset{r\rightarrow\infty}{\lim}\mathcal{R}_{l}(r)\propto\cos\delta_{l}j_{l}(kr)-\sin\delta_{l}n_{l}(kr),
\ee
where $j_l$ represents the spherical Bessel function and $n_l$ represents the spherical Neumann function. Using $\delta_l$, the transfer cross-section is expressed as 
\be\label{eq8}
\frac{\sigma_{T}k^{2}}{4\pi}=\sum_{l=0}^{\infty}(l+1)\sin^{2}(\delta_{l+1}-\delta_{l}).
\ee
The transfer cross-section $\sigma_T$ provides a quantitative measure of the self-interaction rate of DM particles. By solving the Schr\"odinger equation using the partial wave method and determining the phase shifts, we can obtain $\sigma_T$ for puffy DM. Once the transfer cross-section is known, simulations can be performed to study the impact of self-interactions on the small-scale structure of DM halos. In particular, simulations using a self-interaction cross-section of $\sigma_T/m_{\chi}\sim 1~{\rm cm^2/g}$ have successfully produced DM halos that match the observed features of dwarf galaxies and galaxy clusters. 

\begin{figure}[ht]
	\centering
	\includegraphics[width=7cm]{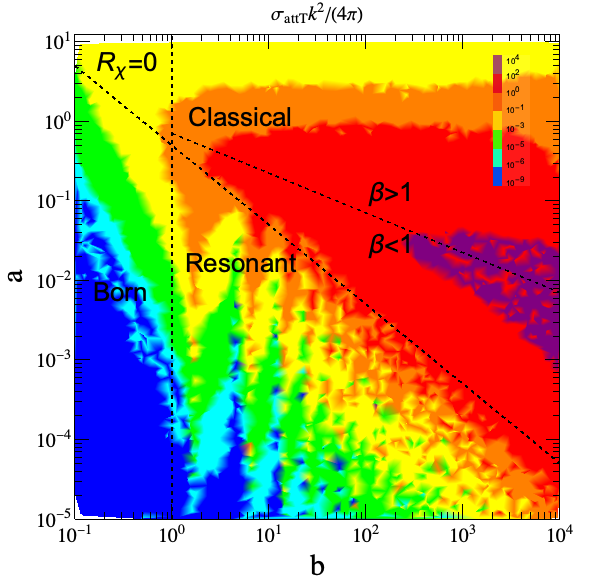}	
	\includegraphics[width=7cm]{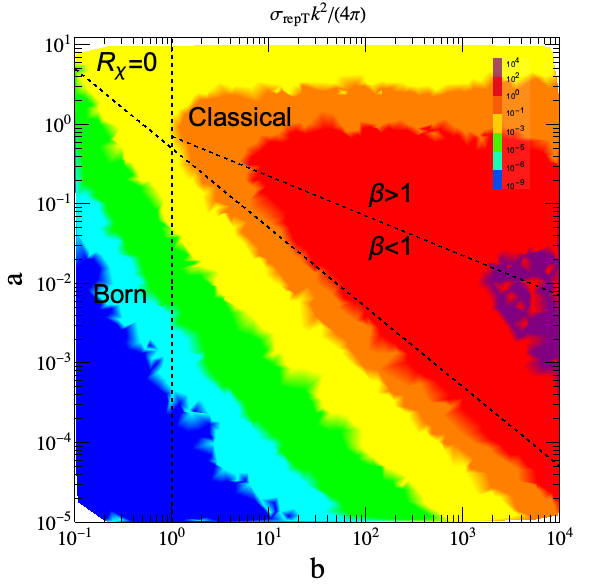}	
	\caption{The parameter space $(a,b)$ for point-like DM, showing the value of $\sigma_Tk^2/(4\pi)$  with attractive (left panel) and repulsive (right panel) force. The Born, classical and resonant parameter regimes are indicated.}
	\label{fig2}
\end{figure}

To study the size effect of puffy SIDM, $\sigma_T$ is obtained by 
solving  the Schrödinger equation with  the new Yukawa potential in
Eq.~(\ref{eq3}). We implement the solution by the software 
Mathematica. The numerical approach has been widely used in particle
physics, including the Sommerfeld enhancement of dark matter annihilation
and self-interacting dark matter. While an approximate solution is
possible for the Hulth\'{e}n potential,
it is not feasible for the puffy Yukawa
potential. To increase the precision of our numerical calculation, we define dimensionless parameters:
\be \label{eq9}
\chi_l=rR_l, \quad  x=\alpha m_{\chi}r, \quad  a=\frac{v}{2\alpha},\quad  b=\frac{\alpha m_{\chi}}{m_{\phi}}\,. 
\ee
The dimensionless radius $x$ is proportional to the dark matter Bohr
radius $1/\alpha m_{\chi}$, while $a$ is the ratio of the relative
velocity to the dark matter Bohr velocity $\alpha$, and $b$ quantifies the
suppression of the Yukawa force range $1/m_{\phi}$ by the Bohr radius
$1/\alpha m_{\chi}$. These parameters enable comparison of different
scenarios for a given system under identical boundary conditions. 
In terms of them, the Schrödinger equation becomes
\be\label{eq10}
\left(\frac{d^{2}}{dx^{2}}+a^{2}-\frac{l(l+1)}{x^{2}}-\frac{1}{m_{\chi}\alpha^{2}}V(r)\right)\chi_{l}=0\, .
\ee
The details of solving the Schr\"odinger equation are presented in Appendix \ref{appb}~\cite{Khrapak:2004,Khrapak:2003kjw}. Here we just briefly describe the calculation.
The initial condition as $\chi_l(x_i)=1$ and $\chi'_l(x_i)=(l+1)/x_i$ 
are set at a point $x_i$ near the origin. This leads to the angular momentum term dominating the Schrödinger equation, which is then solved within the range $x_i \leq x\leq x_m$, with $x_m$ being the maximum value of $x$ used in the numerical analysis. At the condition of  asymptotic solution  Eq.~(\ref{eq7}) and $x=x_m$, we have 
\be\label{eq11}
\chi_l\propto x e^{i\delta_{l}}(\cos\delta_lj_l(ax)-\sin\delta_ln_l(ax)).
\ee
Then the phase shift can be obtained  by
\be\label{eq12}
\tan\delta_l=\frac{ax_mj'_l(ax_m)-\beta_lj_l(ax_m)}{ax_mn'_l(ax_m)-\beta_ln_l(ax_m)},\quad \beta_l=\frac{x_m\chi'_l(x_m)}{\chi_l(x_m)}-1
\ee
When $R_{\chi}=0$, the $(a,b)$ parameter space is scanned and the resulting values of $\sigma_Tk^2/(4\pi)$ for attractive and repulsive forces are shown in Fig.~\ref{fig2}, which agrees with the results in ~\cite{Tulin:2013teo}. This indicates that for the attractive force the transfer cross section can be classified into Born, resonant or classical regimes, as shown in the left panel of Fig.~\ref{fig2}. Here  the parameter $\beta=2\alpha m_{\phi}/(m_{\chi}v^2)$. For $\beta \ll 1$ the classical scattering exhibits a weak Coulomb-like interaction, while for $\beta \gg 1$ a strong potential is observed.  Note that  the repulsive force does not exhibit any resonant phenomenon, as shown in the right panel of Fig.~\ref{fig2}.

\begin{figure}[ht]
	\centering
	\includegraphics[width=5.09cm]{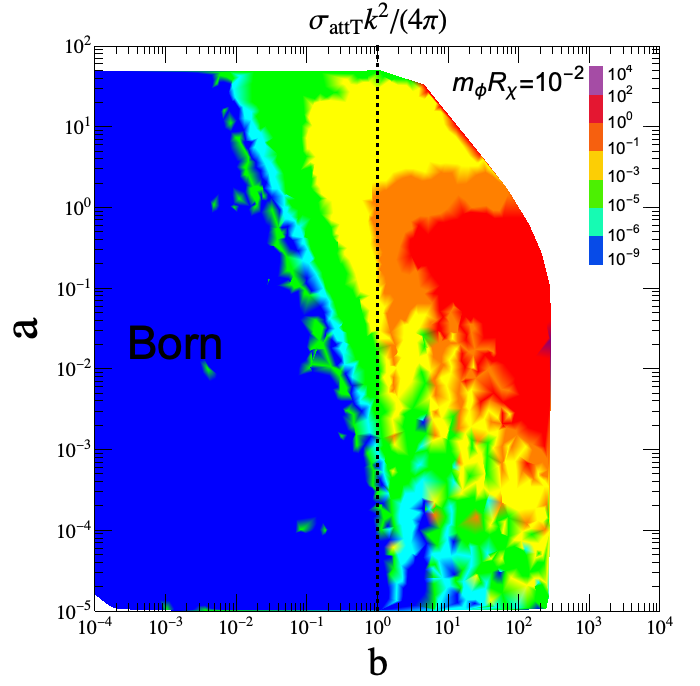}\hspace{-5mm}
	\includegraphics[width=4.8cm]{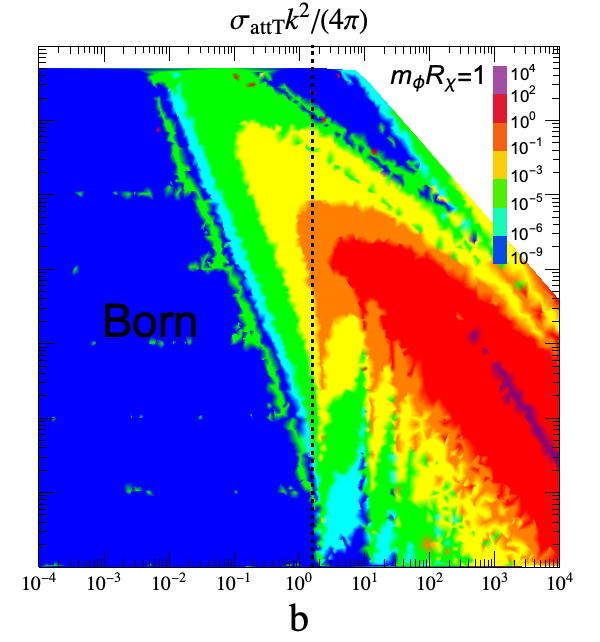}\hspace{-5mm}
	\includegraphics[width=4.8cm]{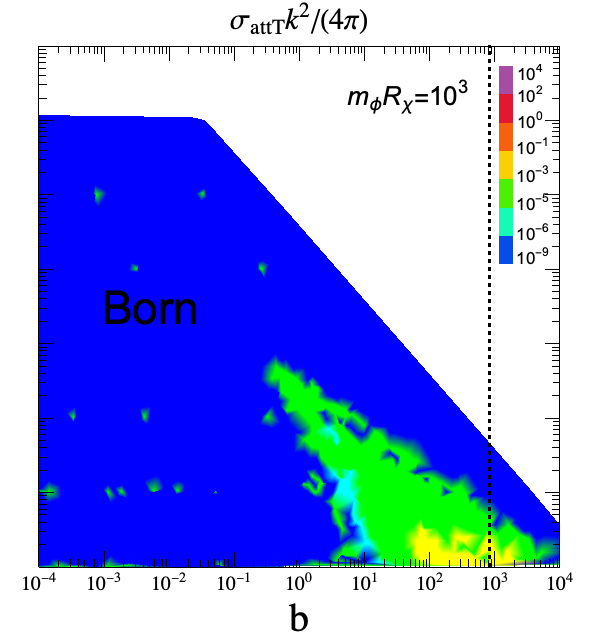}\\	
	\includegraphics[width=5.09cm]{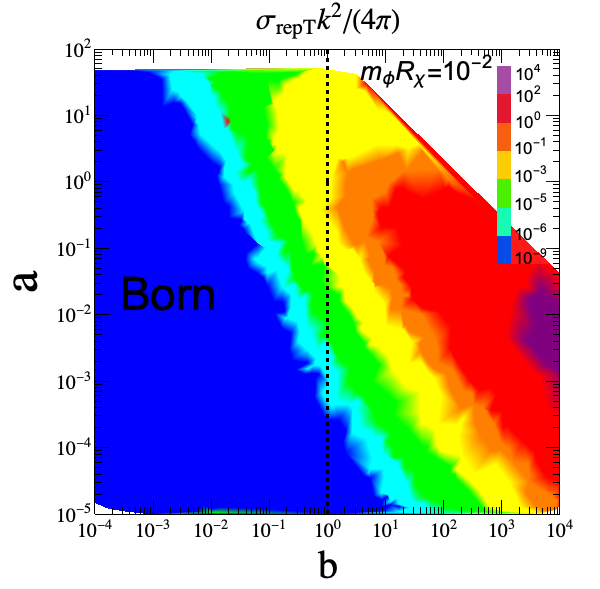}	\hspace{-5mm}
	\includegraphics[width=4.8cm]{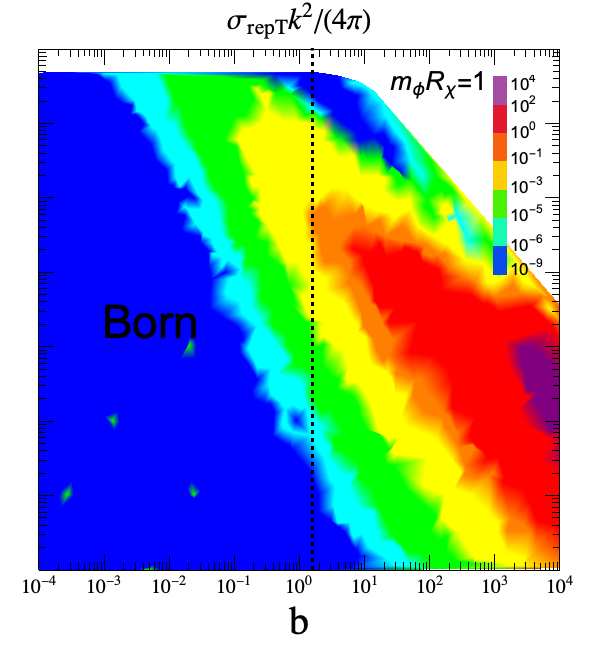}\hspace{-5mm}
	\includegraphics[width=4.8cm]{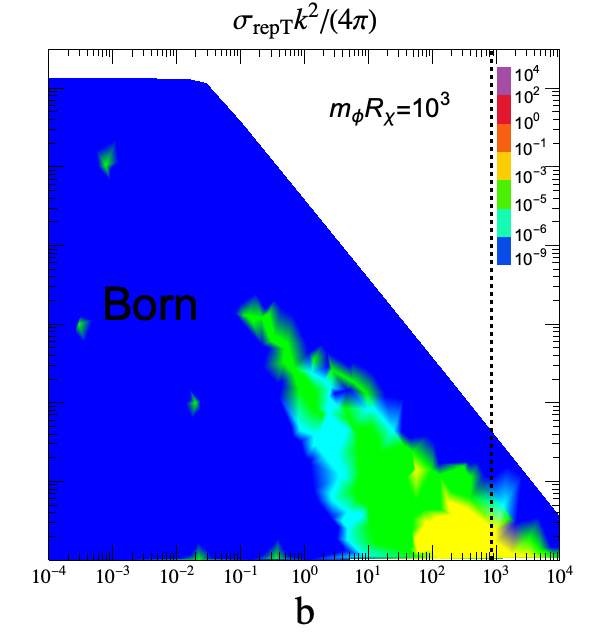}\\	
	\caption{The parameter space $(a,b)$ for puffy DM with $y=$ $10^{-2}$(left panels), $1$ (middle panels), $10^3$ (right panels), showing the value of $\sigma_Tk^2/(4\pi)$  for attractive (upper panels) and repulsive (lower panels) force. 
    The left parts of the vertical line are the valid regions for Born approximation.}
	\label{fig3}
\end{figure}

In  case of puffy DM, the new interacting potential Eq.~(\ref{eq3})
is substituted  to the Schr\"odinger equation in Eq.~(\ref{eq10}).
We vary the constant radius-force range ratio $y$ with values of $10^{-2}, 1, 10^3$ to compare with point DM. By scanning the parameters ($m_{\chi},R_{\chi},\alpha,v,m_{\phi}$), we obtain the $(a,b)$ values and $\sigma_T$ for both attractive and repulsive forces. 
Fig.~\ref{fig3} shows the parameter space $(a, b)$ and the corresponding
value of $\sigma_Tk^2/(4\pi)$ for the different values of $y$.

From Fig.~\ref{fig3} we can see that for small radius-force range ratios $y$, the transfer cross section can still be divided into Born, resonant, or classical regimes for puffy attractive force. 
However, as discussed above, the puffy DM scenario can remove 
the divergent pole of the potential. Thus, even in case of a very small
radius-force range ratio $y$ (the left panels), the transfer cross section $\sigma_T$ is very different from the point particle model. A very obvious feature of the plots is that $\sigma_T$
is almost irrelevant to $b$ in the left Born regimes.
This can be understood from the left panel of  Fig.~\ref{fig1} and the Schr\"odinger equation Eq.~\eqref{eq10}. The transfer cross section $\sigma_T$ is dominated by the parameter
$a$ and the third term of Eq.~\eqref{eq10} since the potential $V_{\rm puffy}(r)$ is small and can be ignored  for $r<2R_{\chi}$ region. For $r>2R_{\chi}$ region, the potential becomes  even  smaller and can also be neglected.

For larger values of $y$, the Born regimes are extended, and the resonant
$\sigma_{{\rm att} T}$ appears for smaller values of $a$. However, for even greater values of $y$, all $V_{\rm puffy}(r)$ will rapidly decrease to zero, and $a^2$ will dominate the cross-section, resulting in no resonant phenomenon, which is similar to the repulsive force case.

In summary, the self-interaction DM cross section in the puffy Yukawa potential is very different from that point DM case. Even though the Born approximation conditions are consistent in the case of extremely small values of $y$, the resonant phenomena and their behavior for the different regions are significantly different.

\section{Implications on the Small-Scale Anomalies}\label{sec4}
\begin{figure}[ht]
	\centering
	\includegraphics[width=5.09cm]{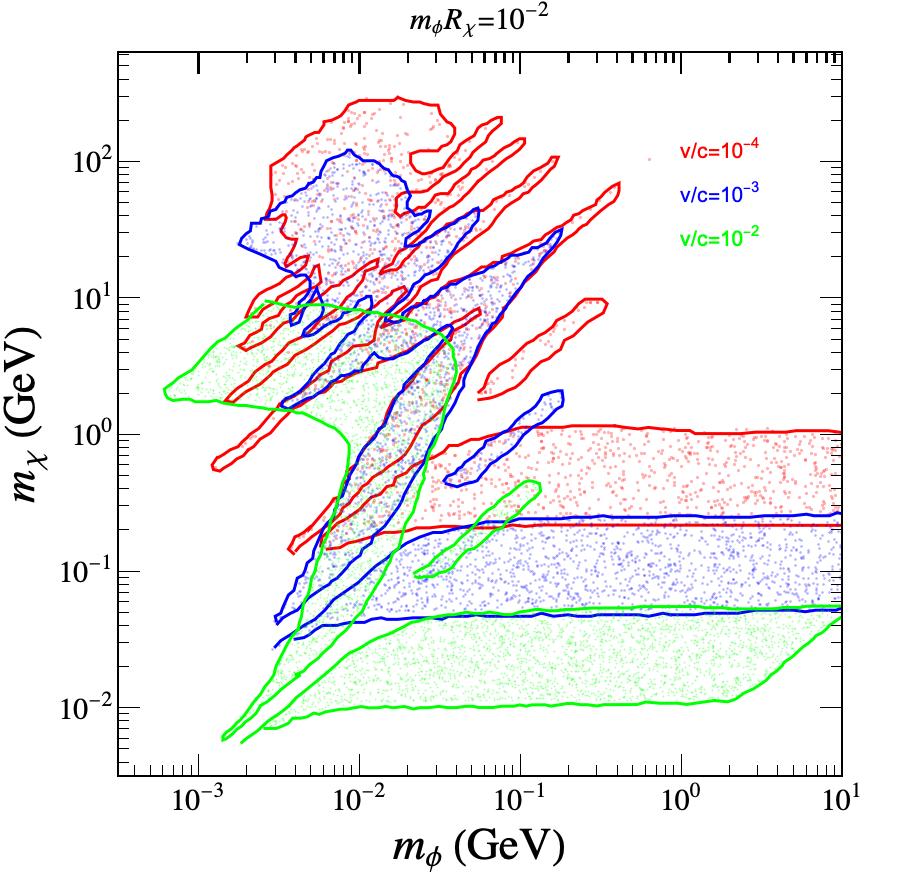}\hspace{-10mm}	
	\includegraphics[width=4.8cm]{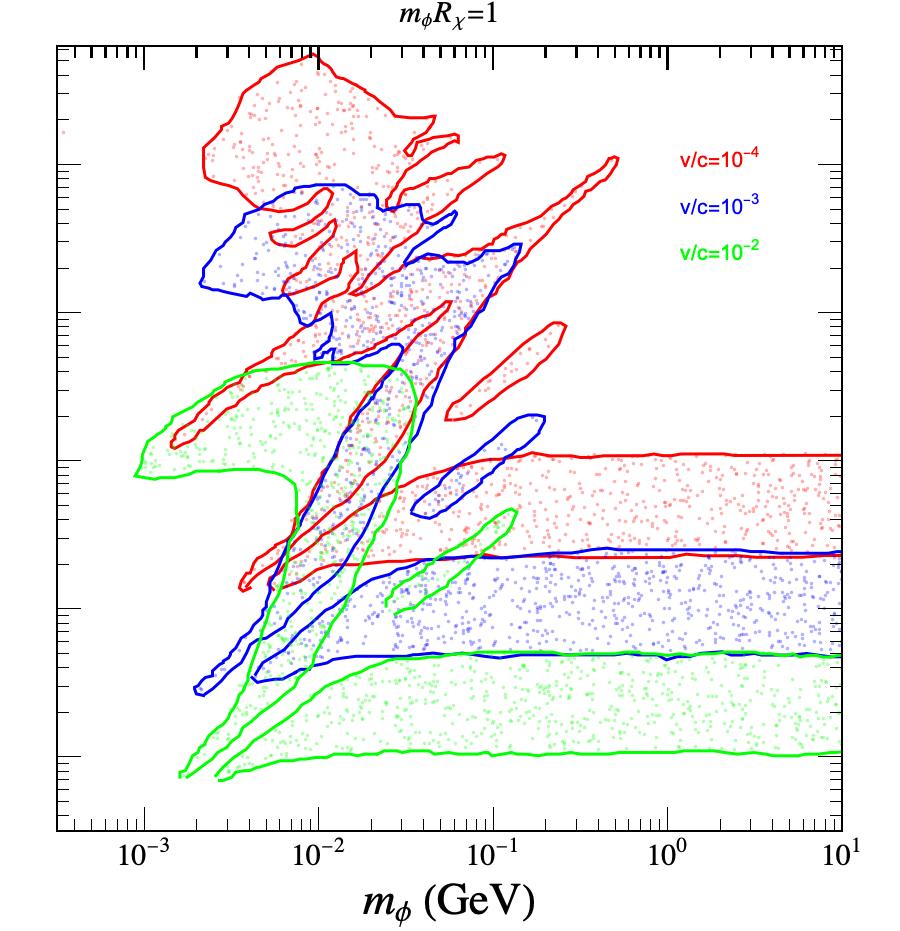}\hspace{-10mm}
	\includegraphics[width=4.8cm]{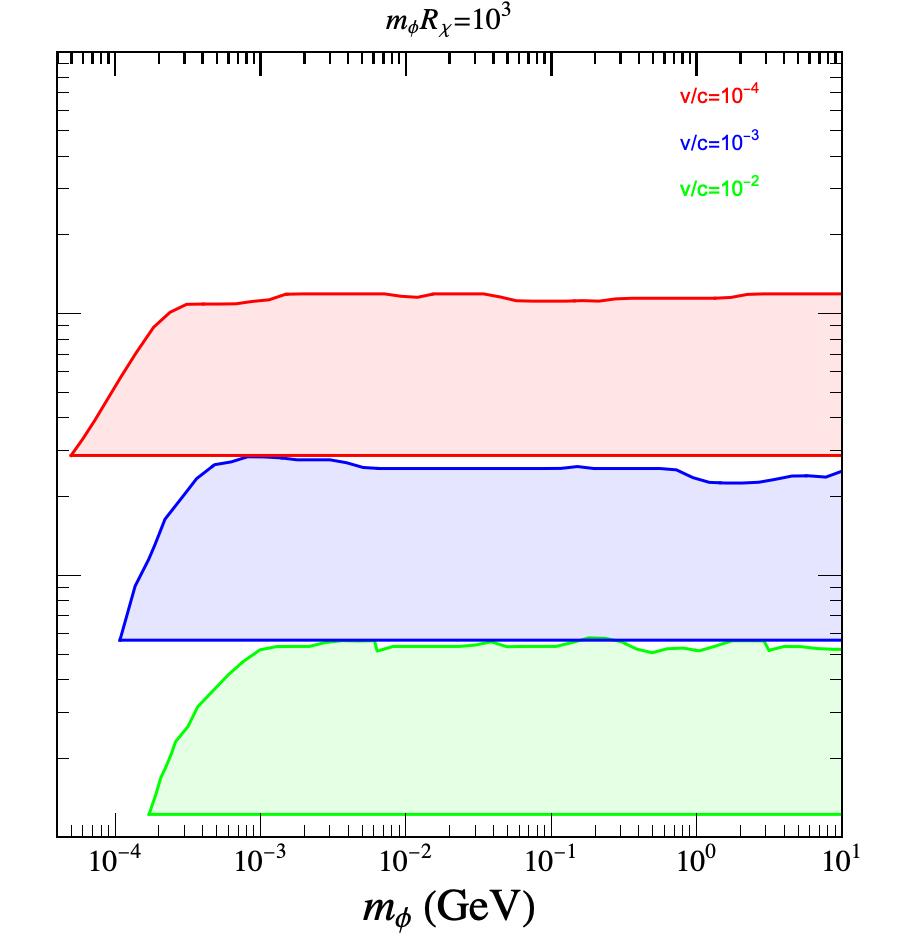}
	\caption{The parameter space for the puffy SIDM self-scattering cross section per unit mass  $0.1~{\rm cm^2/g}<\sigma_T/m_{\chi}<10~ \rm{cm^2/g}$ at different small cosmological scales with a fixed velocity $v/c=10^{-4}$ (red points), $10^{-3}$ (blue points), $10^{-2}$ (green points) corresponding to the dwarf galaxies, Milky Way galaxy and cluster scales, respectively.
 Here $\alpha=10^{-2}$ and the radius-force range ratio $y=10^{-2}$ (left panel), $1$ (middle panel), $10^3$ (right panel) for the  attractive force.}
	\label{fig4}
\end{figure}

In this section we investigate the puffy SIDM model as a potential solution to the small-scale anomalies. We constrain the puffy SIDM self-scattering cross section per unit mass to $0.1~{\rm cm^2/g}<\sigma_T/m_{\chi}<10~ \rm{cm^2/g}$ at different small cosmological scales with a fixed velocity $v/c=10^{-4}, 10^{-3}, 10^{-2}$ corresponding to the dwarf galaxies, Milky Way galaxy and cluster scales, respectively. With $\sigma /m_{\chi}$ in the above range,  we obtain the parameter space for puffy DM scattering in Fig.~\ref{fig4} for the attractive force case. The resonant $\sigma_{{\rm att}T}$ may be produced for a small fixed radius-force range ratio, as shown in the left and middle panels of Fig.~\ref{fig4}, similar to Fig.~\ref{fig3}. However, the resonant effect disappears at high velocities, as shown for $a=v/(2\alpha)\sim 1$ in Fig.~\ref{fig3}, where $\sigma_{{\rm arr}T}$ does not belong to the resonant region. The horizontal areas in Fig.~\ref{fig4} correspond to the Born regimes, where the potential term of the Schr\"odinger equation can be ignored in the parameter space. This is also why the right panel of Fig.~\ref{fig4} appears when $y$ is large. Thus, if the DM has a fixed size, a small $y$ can solve the small cosmological scale anomalies, which is the all-overlap region of Fig.~\ref{fig4} for the attractive force case. 

\begin{figure}[ht]
	\centering
	\includegraphics[width=8.cm]{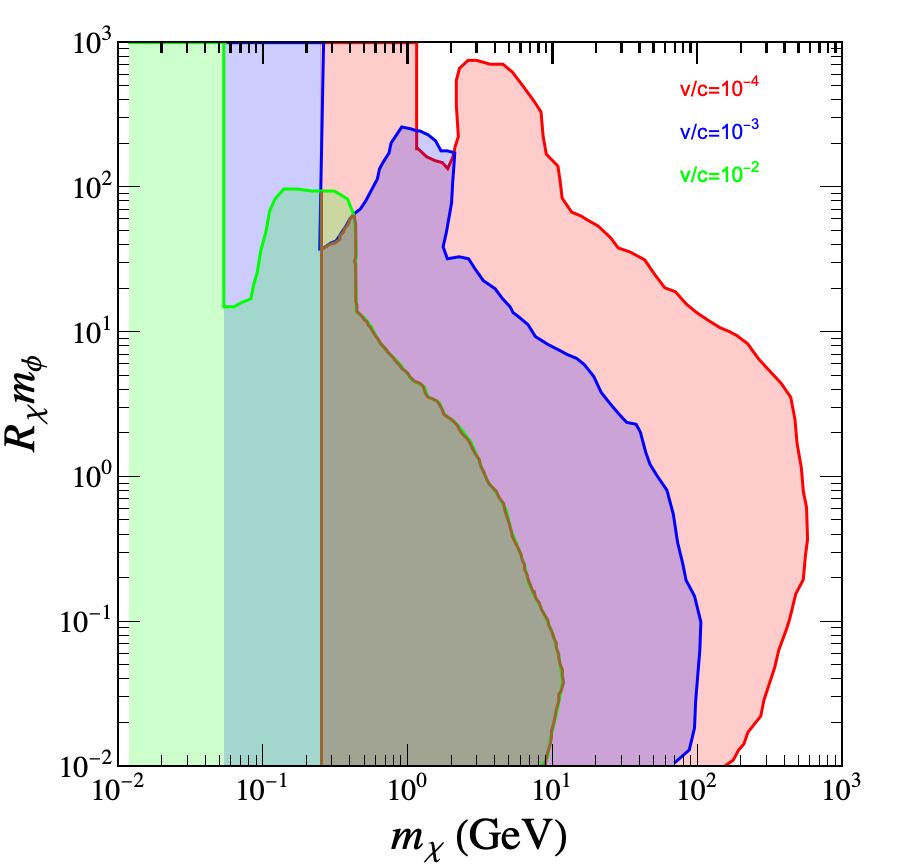}	
	\caption{Same as Fig.~\ref{fig4}, but showing in the plane of ($m_{\chi}$, $y$) with fixed $\alpha=10^{-2}$. Each rectangular region corresponds to the Born regime of puffy DM self-interactions, followed by the resonant and classical regions.}
	\label{fig5}
\end{figure}

We vary the value of $y$ and show in Fig.~\ref{fig5} the parameter space for the puffy SIDM self-scattering cross section solving the small-scale anomalies.  
Here, each rectangular region corresponds to the Born regime of self-interactions, followed by the resonant or classical regime. Notably, the Born regimes become narrower for larger values of $a$, which leads to early appearance of resonant regimes. The overlapping region of Fig.~\ref{fig5} indicates that the small-scale anomalies can be simultaneously addressed.  Specifically, in the parameter space $m_{\chi} \in (230~\rm MeV \sim 10~\rm GeV)$ and $y < 60$,
the puffy SIDM self-scattering can explain the dwarf galaxies in the Born and resonant regions,
and can also explain the cluster and Milky Way galaxy in the non-Born region. So the cross sections merely in the Born approximation can not solve the small-scale anomalies.

\begin{figure}[ht]
	\centering
	\includegraphics[width=8.cm]{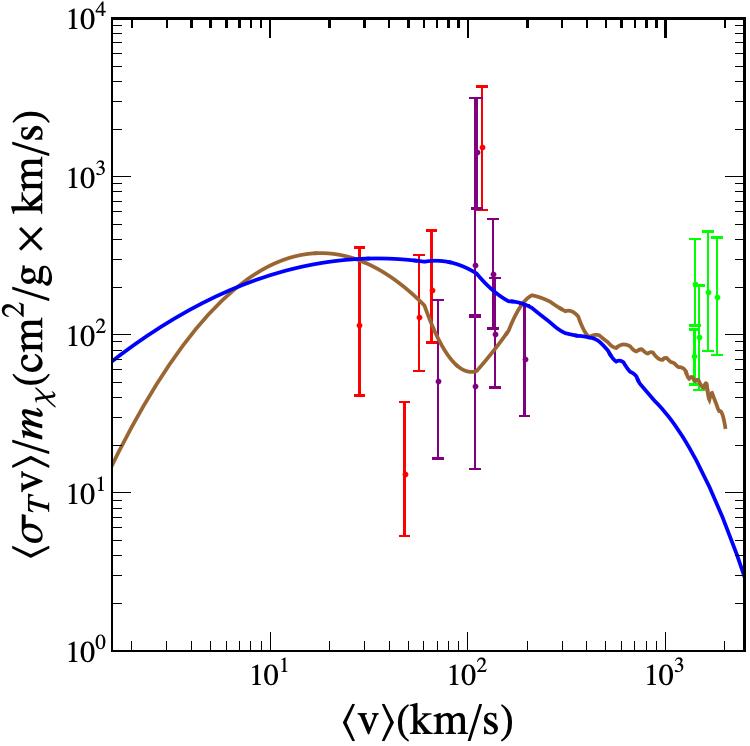}	
	\caption{The transfer cross section of puffy DM versus the velocity for blue line ( $R_{\chi}=80 \rm~ GeV^{-1}, m_{\phi}= 0.01\rm ~GeV,\alpha=10^{-2}$ and $m_{\chi}=15~\rm GeV$) and brown line($R_{\chi}=30~ \rm GeV^{-1}, m_{\phi}= 0.1~\rm GeV,\alpha=10^{-1}$ and $m_{\chi}=10~\rm GeV$). The data points corresponding to the dwarfs, the LSB galaxies and the clusters are shown in red, purple, and green, respectively.} 
	\label{fig6}
\end{figure}

In the astrophysical halos, the DM follows a Maxwell-Boltzmann distribution. The velocity-averaged transfer cross section is 
\bea \label{vat}
\langle \sigma_{T}v \rangle=\int f(v)\sigma_T vdv, 
\eea
with 
\bea
f(v)=\frac{32v^2e^{-4v^2/\pi\langle v\rangle^2}}{\pi^2\langle v\rangle^3} .
\eea 
The velocity $v$ refers to the relative velocity in the center-of-mass frame. 
The puffy SIDM benchmark point with $R_{\chi}=30\rm GeV^{-1}$, $m_{\phi}=0.1\rm GeV$, $\alpha=10^{-1}$ and $m_{\chi}=10\rm GeV$ gives the results in Fig.~\ref{fig6}, which are compared with the observed data.
So we see that the puffy SIDM explain the data quite well. Beyond that, as in the point-like DM case, the velocity-dependence of $\sigma_T$ usually has a small turn-over that can be understood as going from the constant scattering to Rutherford-like scattering.  In the puffy DM case, this can happen for some parameters, such as the blue curve in Fig.~\ref{fig6}. And as shown by the numerical study of point DM scattering in \cite{Tulin:2013teo},
the most important feature is the highly nontrivial velocity-dependence of $\sigma_T$ in the resonant region. In the puffy DM case, for some values of $y$ and $b$, the resonance or anti-resonance feature can also occur, as shown by the brown curve in Fig.~\ref{fig6}.

Finally, we expect that the radius-force range ratio $y$ to be constrained by the direct detections of puffy DM, as explored in some recent works \cite{Monteiro:2020wcb, Acevedo:2020avd,Laha:2015yoa}.
Also, in our future work we expect to examine the puffy DM form factor and provide more precise solutions to the Schrodinger equation. 
 
\section{Conclusion}\label{sec5}

In this work, we investigated the puffy dark matter with a tophat form factor, considering the piecewise puffy Yukawa potential. Our results showed that the potential rapidly decreases with $y$ for $r>2R_{\chi}$, which is different from the usual Yukawa potential. Furthermore, the Born effective regime is expanded in the puffy Yukawa potential case compared to the point particle case. We revisited the self-interacting DM scattering by solving the Schrödinger equation using the partial wave approach. Our analysis revealed that for small values of $y$, the potential term can be ignored due to the radius effect, and the resonant regime scattering can solve the small cosmological scale anomalies. However, for larger $y$ values, the scattering is almost non-existent, and the small-scale problems may not be solved. Our investigation showed that the self-interacting puffy DM can explain the dwarf galaxies in the Born and resonant regimes, and also can explain the cluster and Milky Way galaxy in 
the non-Born regime.  Thus, the puffy DM self-scattering in the Born approximation alone cannot explain the small-scale anomalies.
 
\section*{Acknowledgements}
This work was supported by the Natural Science Foundation of China (NSFC) under grant numbers 12075300, 11821505, and 12275232, the Peng-Huan-Wu Theoretical Physics Innovation Center (12047503), the CAS Center for Excellence in Particle Physics (CCEPP), the Key $R\&D$ Program of Ministry of Science and Technology of China under number 2017YFA0402204, and the Key Research Program of the Chinese Academy of Sciences under grant No. XDPB15.

\appendix
\section{The puffy Yukawa potential}\label{appa}
Here we deduce the puffy Yuawa potential.  First, when the interaction belongs to Yukawa force among two-point particles, the potential can be written as 
\be \label{Apot}
V_{00}=\frac{Q^2e^{-m_{\phi r}}}{4\pi r}.
\ee 
\begin{figure}[ht]
	\centering
	\includegraphics[width=10.cm]{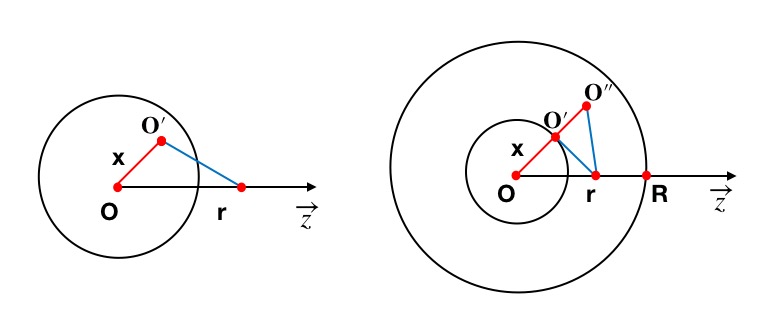}	
	\caption{On the left, the point particle being tested, which possesses a dark charge Q, is located outside the puffy dark matter sphere. On the right, the test point particle is located within the puffy dark matter sphere.}
	\label{fig7}
\end{figure}
Thus, for a homogenous sphere with a dark charge density of $3Q/(4\pi R_{\chi}^3)$, an analogous field to the electric field in space is generated. When the test point particle with a dark charge Q is located outside of the puffy dark matter ball, as shown in the left image of Fig.~\ref{fig6}, the potential outside of this point-ball system can be treated as the integration of all point charge potentials for this sphere. Its expression is
\be \label{poi}
\begin{split}
V_{0}(r)&=\int\frac{3Q}{4\pi R^{3}}V_{00}dV\\
&=-\frac{3Q}{4\pi R^{3}}\int^R_0\frac{Q}{4\pi\sqrt{x^{2}+r^{2}-2xr\cos\theta}}e^{-m_{\phi}\sqrt{x^{2}+r^{2}-2xr\cos\theta}}x^{2}dxd\cos\theta2\pi\\
&=\frac{3Q}{4\pi R^{3}}\frac{Qe^{-m_{\phi}r}}{2m_{\phi}r}\left(\frac{e^{-m_{\phi}R}}{m_{\phi}^{2}}-\frac{e^{m_{\phi}R}}{m_{\phi}^{2}}+\frac{Re^{-m_{\phi}R}}{m_{\phi}}+\frac{Re^{m_{\phi}R}}{m_{\phi}}\right).
\end{split}
\ee 
At a point inside the sphere, the Yukawa potential is from the contribution of the charge of   the ball with radius r and the spherical shell $(r,R)$. In the right picture of Fig.~\ref{fig6},  when  $0<x<r$,  the  potential can be obtained as 
\be \label{inbol1}
\begin{split}
	V_{1}(r)&=-\int_{0}^{r}\frac{3Q}{4\pi R_{\chi}^{3}}V_{00}x^{2}dxd\cos\theta2\pi\\
	&=-\frac{3Q}{4\pi R^{3}}\frac{Q}{2m_{\phi}r}\left.\left(-\frac{e^{-m_{\phi}(r+x)}(1+m_{\phi}x+e^{2m_{\phi}x}(m_{\phi}x-1))}{m_{\phi}^{2}}\right)\right|_{0}^{r}\\
	&=\frac{3Q}{4\pi R^{3}}\frac{Q}{2m_{\phi}r}\frac{e^{-2m_{\phi}r}(1+m_{\phi}r)+(m_{\phi}r-1)}{m_{\phi}^{2}}.
\end{split}
\ee
When  $r<x<R$,  the  potential  is 
\begin{eqnarray}
 \label{inbol2}
	V_{2}(r)&=&-\int_{r}^{R}\frac{3Q}{4\pi R_{\chi}^{3}}V_{00}x^{2}dxd\cos\theta2\pi \nonumber\\
	&=&-\frac{3Q}{4\pi R_{\chi}^{3}}\frac{Q}{2m_{\phi}r}\left.\left(-\frac{e^{-m_{\phi}(r+x)}(1+m_{\phi}x+e^{2m_{\phi}x}(m_{\phi}x-1))}{m_{\phi}^{2}}\right)\right|_{r}^{R_{\chi}}\nonumber\\
	&=&\frac{3Q}{4\pi R_{\chi}^{3}}\frac{Q}{2m_{\phi}r}\left(\frac{(e^{-m_{\phi}(r+R_{\chi})}-e^{-m_{\phi}(R_{\chi}-r)})(1+y)}{m_{\phi}^{2}} \right.\nonumber \\
&& \left. -\frac{e^{-2m_{\phi}r}+m_{\phi}re^{-2m_{\phi}r}-1-m_{\phi}r}{m_{\phi}^{2}}\right).
\end{eqnarray}
Thus,  inside the sphere, the total potential  is 
\be\label{ins}
V_{3}=-\frac{3Q}{4\pi R_{\chi}^{3}}\frac{Q}{2m_{\phi}r}\frac{e^{-m_{\phi}(r+R_{\chi})}(e^{2m_{\phi}r}-1)(1+y)}{m_{\phi}^{2}}+\frac{3Q}{4\pi R_{\chi}^{3}}\frac{Q}{m_{\phi}^{2}}.
\ee 
Accordingly, in the point-ball system, the Yukawa potential in the space can be expressed as 
\small
\begin{align}\label{bigr}
V_4(r)&=\begin{cases}
-\frac{3Q}{4\pi R_{\chi}^{3}}\frac{Q}{2m_{\phi}r}\frac{e^{-m_{\phi}(r+R_{\chi})}(e^{2m_{\phi}r}-1)(1+y)}{m_{\phi}^{2}}+\frac{3Q}{4\pi R_{\chi}^{3}}\frac{Q}{m_{\phi}^{2}}\  & \left(0<r<R_{\chi}\right)\\
\hspace{2cm}\  & \ \\[-4.1mm]
\frac{3Q}{4\pi R_{\chi}^{3}}\frac{Qe^{-m_{\phi}r}}{2m_{\phi}r}\left(\frac{e^{-y}}{m_{\phi}^{2}}-\frac{e^{y}}{m_{\phi}^{2}}+\frac{R_{\chi}e^{-y}}{m_{\phi}}+\frac{R_{\chi}e^{y}}{m_{\phi}}\right) & \left(R_{\chi}<r\right). 
\end{cases}
\end{align}
\normalsize
When the charge of the field source is from a sphere and the test particle is also a sphere. Out of the sphere,  the potential can be gotten via integrating the  all point charge from the test sphere as the left picture of Fig.~\ref{fig7} and it is 
\small
\be \label{obol1}
\begin{split}
	V_{5}(r)&=(\frac{3Q}{4\pi R_{\chi}^{3}})^{2}\int_{0}^{\sqrt{R_{\chi}^{2}-(r-x)^{2}}}V_{0}(r=\sqrt{x^{2}+y^{2}})2\pi ydy\int_{r-R_{\chi}}^{r+R_{\chi}}dx\\
	&=\left(\frac{3Q}{4\pi R_{\chi}^{3}}\right)^{2}\frac{e^{-m_{\phi}r}}{m_{\phi}^{2}r}\pi\left(\frac{e^{-y}}{m_{\phi}^{2}}-\frac{e^{y}}{m_{\phi}^{2}}+\frac{R_{\chi}e^{-y}}{m_{\phi}}+\frac{R_{\chi}e^{y}}{m_{\phi}}\right)^{2}.
\end{split}
\ee
\normalsize
\begin{figure}[ht]
	\centering
	\includegraphics[width=7.cm]{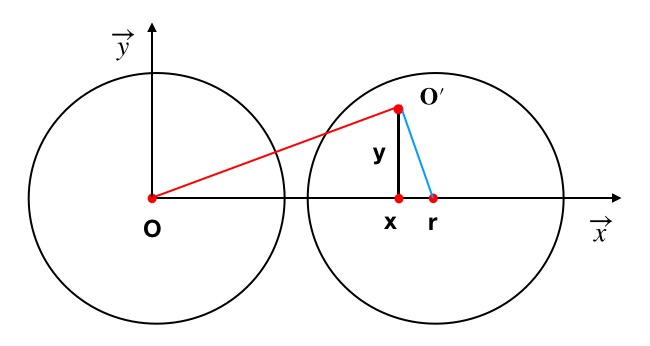}	
	\includegraphics[width=7.cm]{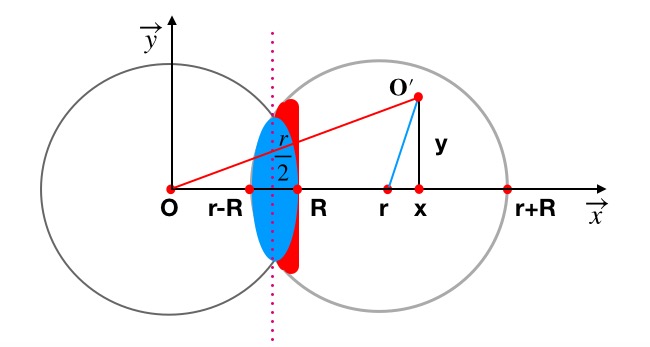}
	\caption{Left: the test ball with dark charge Q is outside of the puffy dark matter ball. Right: the test ball with dark charge Q is inside of the  puffy dark matter ball.}
	\label{fig8}
\end{figure}
When $R_{\chi}<r<2R_{\chi}$,  the potential  can be devided into four part.
In the region $(r-R_{\chi}<r<r/2)$ and  $(r/2<r<R_{\chi})$ of test sphere, namely, the left half of the blue area of right picture of  Fig.~\ref{fig7}  and the right  that of  Fig.~\ref{fig7}, the potential can be gotten via the $V_3(r)$.  Because these test charge is inside the source charge space.  In the region $(r/2<r<R_{\chi})$ of the red  area of right picture of  Fig.~\ref{fig7}  and the region $(R_{\chi}<r<r+R_{\chi})$, the potential can be obtained by the $V_0(r)$.
Thus, the potential  expression can be written as 
\begin{align}\label{iigr}
V_6(r)&=\begin{cases}
\frac{3Q}{4\pi R_{\chi}^{3}}\int_{r-R_{\chi}}^{\frac{r}{2}}\int_{0}^{\sqrt{R_{\chi}^{2}-(r-x)^{2}}}V_{3}(r=\sqrt{x^{2}+y^{2}})2\pi ydydx\  & \left(r-R_{\chi}<r<r/2\right)\\
\hspace{2cm}\  & \ \\[-4.1mm]
\frac{3Q}{4\pi R_{\chi}^{3}}\int_{\frac{r}{2}}^{R_{\chi}}\int_{0}^{\sqrt{R_{\chi}^{2}-x^{2}}}V_{3}(r=\sqrt{x^{2}+y^{2}})2\pi ydydx & \left(r/2<r<R_{\chi}\right)_{\rm blue}\\
\hspace{2cm}\  & \ \\[-4.1mm]
\frac{3Q}{4\pi R_{\chi}^{3}}\int_{\sqrt{R_{\chi}^{2}-x^{2}}}^{\sqrt{R_{\chi}^{2}-(r-x)^{2}}}V_{0}(r=\sqrt{x^{2}+y^{2}})2\pi ydy\int_{\frac{r}{2}}^{R_{\chi}}dx & \left(r/2<r<R_{\chi}\right)_{\rm red}\\
\hspace{2cm}\  & \ \\[-4.1mm]
\frac{3Q}{4\pi R_{\chi}^{3}}\int_{0}^{\sqrt{R_{\chi}^{2}-(r-x)^{2}}}V_{0}(r=\sqrt{x^{2}+y^{2}})2\pi ydy\int_{R_{\chi}}^{r+R_{\chi}}dx& \left(R_{\chi}<r<r+R_{\chi}\right)\\
\hspace{2cm}\  & \ \\[-4.1mm]
\end{cases}
\end{align}
When  $0<r<R_{\chi}$, the way of calculating the  potential  is same as the $R_{\chi}<r<2 R_{\chi}$ case and the result is also identical.
Thus, when $0<r<2R_{\chi}$,  the total potential is 
\be \label{ibol1}
\begin{split}
V_{7}(r)=&-\alpha4\pi^{2}(\frac{3}{4\pi R_{\chi}^{3}})^{2}\frac{(1+y)e^{-y}}{m_{\phi}^{3}}\frac{1}{2m_{\phi}^{3}}\times\\
&\left(\frac{2e^{-m_{\phi}(R_{\chi}+r)}(-1+e^{y})(e^{2y}(-1+y)+e^{y}(1+y))}{r}\right.\\
&\Bigg.+m_{\phi}(-e^{y}(2+m_{\phi}(r-2R_{\chi}))+e^{-y}(2-m_{\phi}r+2y))\Bigg)\\
&+\alpha4\pi^{2}\left(\frac{3}{4\pi R_{\chi}^{3}}\right)^{2}\frac{1}{m_{\phi}}\left(\frac{e^{-y}}{m_{\phi}^{2}}-\frac{e^{y}}{m_{\phi}^{2}}+\frac{R_{\chi}e^{-y}}{m_{\phi}}+\frac{R_{\chi}e^{y}}{m_{\phi}}\right)\\
&\frac{e^{-m_{\phi}(R_{\chi}+r)}(2+2y-e^{y}(2+m_{\phi}r(-2+m_{\phi}(r-2R_{\chi}))+2y))}{2m_{\phi}^{3}r}\\
&+\alpha4\pi^{2}\left(\frac{3}{4\pi R_{\chi}^{3}}\right)^{2}\frac{1}{m_{\phi}^2}\left( \frac{1}{12}(r-2R_{\chi})(r+4R_{\chi})\right), 
\end{split}
\ee
where $\alpha=Q^2/4\pi$.  Accordingly, the Yukawa potential  among two sphere is 
\begin{align}\label{eqa}
V_{\rm puffy}(r) & = \begin{cases}
\pm g(y) & r<2R_{\chi} \\
\hspace{5cm}\  & \ \\[-6.mm]
\pm\alpha\frac{e^{-m_{\phi}r}}{r} \times h\left(y\right)&  r>2R_{\chi},
\end{cases}
\end{align}
where
\be \label{inbol}
\begin{split}
g(r,y)=&-\alpha4\pi^{2}(\frac{3}{4\pi R_{\chi}^{3}})^{2}\frac{(1+y)e^{-y}}{m_{\phi^{3}}}\frac{1}{2m_{\phi}^{3}}\\
&\times\left(\frac{2e^{-m_{\phi}(R_{\chi}+r)}(-1+e^{y})(e^{2y}(-1+y)+e^{y}(1+y))}{r}\right.\\
&\Bigg.+m_{\phi}(-e^{y}(2+m_{\phi}(r-2R_{\chi}))+e^{-y}(2-m_{\phi}r+2y))\Bigg)\\
&+\alpha4\pi^{2}\left(\frac{3}{4\pi R_{\chi}^{3}}\right)^{2}\frac{1}{m_{\phi}}\left(\frac{e^{-y}}{m_{\phi}^{2}}-\frac{e^{y}}{m_{\phi}^{2}}+\frac{R_{\chi}e^{-y}}{m_{\phi}}+\frac{R_{\chi}e^{y}}{m_{\phi}}\right)\\
&\times\frac{e^{-m_{\phi}(R_{\chi}+r)}(2+2y-e^{y}(2+m_{\phi}r(-2+m_{\phi}(r-2R_{\chi}))+2y))}{2m_{\phi}^{3}r}\\
&+\alpha4\pi^{2}\left(\frac{3}{4\pi R_{\chi}^{3}}\right)^{2}\frac{1}{m_{\phi}^2}\left( \frac{1}{12}(r-2R_{\chi})(r+4R_{\chi})\right), 
\end{split}
\ee
\be \label{outboll}
\begin{split}
h\left(y\right)&=4\pi\left(\frac{3}{4\pi R_{\chi}^{3}}\right)^{2}\frac{\pi}{m_{\phi}^{2}}\left(\frac{e^{-y}}{m_{\phi}^{2}}-\frac{e^{y}}{m_{\phi}^{2}}+\frac{R_{\chi}e^{-y}}{m_{\phi}}+\frac{R_{\chi}e^{y}}{m_{\phi}}\right)^{2}
\end{split}
\ee
\section{The details with \texorpdfstring{$x_m$}{}  and \texorpdfstring{$\sigma_T$}{}}\label{appb}
To solve Eq.~(\ref{eq10}),  the value of $x_m$ is given via the condition $a^2\ll exp(-x_m/b)/x_m$.  So the potential term can be ignored compared to the kinetic term. In fact, when  $x_m$ is very large,   Eq.~(\ref{eq10}) is dominated by the kinetic term  and $\delta_l$ can be obtained.  In our  Mathematical method, when $l=0$,  if $a\geq1, x_m =10^3$, or $x_m=50/(10^{c-1})$ for point-like DM (if puffy DM case, $x_m=10/(10^{c-3})$)  and c is the integer part of $\rm Log10(a)$.  For  range  $1\leq l<5$, $x_m=\sqrt{l+l^2}\sqrt{3\times 10^6}/a $. When $5\leq l <10$, $x_m=\sqrt{l+l^2}\sqrt{3\times 10^3}/a $ and  $10 \leq l \leq 15$, $x_m=\sqrt{l+l^2}\sqrt{\times 10^3}/a $.  And in this way $l_{\rm max}$ is imposed as 15. Then the phase shift can be obtained by Eq.~(\ref{eq12}).  

The  transfer cross section $\sigma_T$ is calculated by Eq.~(\ref{eq8}) summing all $l$ until the successive $\sigma_T$ converges  to $0.01$. Note that in our method, if $l>l_{\rm max}$, this  set of parameter data is neglected.


\end{document}